\documentclass[prd,twocolumn]{revtex4-1}

\usepackage{amsmath}
\usepackage{blkarray}
\usepackage{multirow}
\usepackage{mathtools}
\usepackage{graphicx}

\usepackage{graphics}
\usepackage{graphicx}
\usepackage{epsfig}
\usepackage{color}
\begin{document}

\newcommand{\BayesWave}{\texttt{BayesWave}}
\newcommand{\BayesLine}{\texttt{BayesLine}}
\newcommand{\LALInference}{\texttt{LALInference}}
\newcommand{\cWB}{\texttt{cWB}}
\newcommand{\Msun}{{\mathrm{M}_\odot}}
\newcommand{\Mc}{{\mathcal{M}}}
\newcommand{\data}{d}
\newcommand{\h}{{\bm{h}}}
\newcommand{\n}{{\bm{n}}}
\newcommand{\params}{{\boldsymbol \theta}}
\newcommand{\intp}{{\boldsymbol \lambda}}
\newcommand{\extp}{{\boldsymbol \Omega}}
\newcommand{\Hyp}{{\mathcal{H}}}
\newcommand{\Bay}{{\mathcal{B}}}
\newcommand{\Sig}{{\mathcal{S}}}
\newcommand{\Gli}{{\mathcal{G}}}
\newcommand{\Odd}{{\mathcal{O}}}
\newcommand{\IFO}{{\rm IFO}}

\newcommand{\Baysg}{{\mathcal{B_{\Sig , \Gli}}}}

\newcommand\Neil[1]{\textcolor{red}{[#1]}}
\newcommand\Tyson[1]{\textcolor{blue}{[#1]}}
\newcommand\Jonah[1]{\textcolor{green}{[#1]}}

\DeclareGraphicsExtensions{.pdf,.gif,.jpg}

\title{Leveraging waveform complexity for confident detection of gravitational waves}

\author{Jonah B. Kanner}
\affiliation{LIGO Laboratory, California Institute of Technology, Pasadena, CA 91125, USA}
\email{jkanner@caltech.edu}

\author{Tyson B. Littenberg}
\affiliation{Center for Interdisciplinary Exploration and Research in Astrophysics (CIERA) \& Department of Physics and Astronomy, Northwestern University, 2145 Sheridan Road, Evanston, IL 60208}

\author{Neil Cornish}
\affiliation{Montana State University, Bozeman, Montana 59717, USA}

\author{Enia Xhakaj}
\affiliation{Lafayette College, 730 High St, Easton, PA 18042}

\author{Meg Millhouse}
\affiliation{Montana State University, Bozeman, Montana 59717, USA}

\author{Francesco Salemi}
\affiliation{Data Analysis Group, Albert-Einstein-Institut, Max-Planck-Institut für, Gravitationsphysik, D-30167 Hannover, Germany}

\author{Marco Drago}
\affiliation{Data Analysis Group, Albert-Einstein-Institut, Max-Planck-Institut für, Gravitationsphysik, D-30167 Hannover, Germany}

\author{Gabriele Vedovato}
\affiliation{INFN Padova, Via Marzolo 8, Padova I-35131, Italy}

\author{Sergey Klimenko}
\affiliation{University of Florida, Gainesville, FL 32611, USA}


\begin{abstract}
The recent completion of Advanced LIGO suggests that gravitational waves (GWs)
may soon be directly observed.  Past searches for gravitational-wave transients
have been impacted by transient noise artifacts, known as glitches, introduced into 
LIGO data due to instrumental and environmental effects.  In this work, we explore how waveform
complexity, instead of signal-to-noise ratio, can be used to rank event candidates and 
distinguish short duration astrophysical signals from glitches.  
We test this framework using a new hierarchical pipeline 
that directly compares the Bayesian evidence of explicit signal and glitch models.  
The hierarchical pipeline is shown to have strong performance, and in particular,
allows high-confidence detections of a range of waveforms at realistic 
signal-to-noise ratio with a two detector network.


\end{abstract}

\maketitle

\section{Introduction}


Installation of the Advanced LIGO \cite{aLIGO} gravitational wave (GW) detectors has recently been 
completed, and the first observation run began in September of this year.
These new detectors are designed to make detection of GWs 
a reality.  For example, current estimates predict that 
Advanced LIGO will eventually detect between 1 and 400 binary neutron star mergers 
per year \cite{rates_lsc}.  A number of other sensitive GW detectors are in various stages of construction
and installation, including Advanced Virgo \cite{AdVirgo}, GEO600 \cite{GEOHF}, and Kagra \cite{KAGRA}.

The principal analysis challenge in finding transient signals in LIGO 
data is separating signatures of astrophysical sources from large populations 
of transient detector artifacts (glitches) in the data.
Researchers have demonstrated a variety of search techniques for 
finding transient signals with initial LIGO and initial Virgo.  Searches for 
specific classes of signals, including binary neutron 
star mergers and mergers of solar mass black holes, have demonstrated 
performance in real LIGO data similar to expectations based on 
Gaussian noise, suggesting they are optimally sensitive \cite{s6cbc}.
These searches use a matched filtering technique to reject glitches, an approach which relies
on detailed knowledge of the expected waveform.  

On the other hand, LIGO data is also searched for generic GW transients, known as GW bursts, which are not constrained by a specific source model.  Such searches are designed
to detect unmodeled and/or unexpected GW sources.
These searches are less constrained by waveform 
morphology, and so are more sensitive to glitches which diminish detection confidence of potential GW event candidates. 
For example, past searches for bursts in 
LIGO data have shown background distributions that included high signal-to-noise ratio glitches \cite{s5burst, s6burst} which diminish detection confidence of any potential GW events.  

One approach to confident Burst detection is to carefully divide the parameter space, 
and improve detection confidence for particular classes of signals.  
Recent work by Thrane and Coughlin \cite{thrane2015} has shown that searches for
long duration bursts, with time-scales greater than several seconds,
are insensitive to most glitch populations, and so can successfully 
identify long bursts with high confidence.  For short duration searches, it may also be possible
to carefully study background distributions, and apply \textit{ad hoc} cuts designed to 
isolate portions of parameter space that are relatively glitch free.


Taking a different approach, Cornish and Littenberg put forward the \BayesWave{} pipeline \cite{bayeswave}, 
and described how it uses
a novel detection statistic to characterize GW data.  Most Burst 
search algorithms apply selection cuts to remove glitches, and then rank 
the remaining signals with a statistic proportional to 
signal-to-noise ratio (SNR). \BayesWave{} instead attempts to fit the data with both a GW
signal model and an explicit glitch model, and calculate the Bayesian evidence ratio (Bayes Factor)
between the two competing hypotheses.  Because of this, \BayesWave{} may be more robust against the 
high SNR glitches which have been problematic for past searches.

In this work, we describe how \BayesWave{} can be used as a second stage
to follow-up triggers from a leading burst pipeline, \texttt{coherentWaveBurst} (\cWB) \cite{cwb},
to create a 
``hierarchical pipeline'' that combines the best features of the two tools.  
To test the performance, we measure the ability of the joint search to detect 
simulated gravitational waves while rejecting glitches using the two LIGO detectors in Livingston and Hanford. 
We study the performance on 
a range of waveforms, including both binary black hole mergers and several 
\textit{ad hoc} waveforms that have been used in previous burst searches.

\section{The Hierarchical Pipeline}


The \cWB{} pipeline has been used in a number of previous burst searches.  
The algorithm looks for coherent excess power by cross-correlating data streams 
between two or more detector sites, using projection coefficients that reject signal
power which is inconsistent with a source at a hypothesis sky position.  
The detection statistic, $\rho$, is designed to scale with the SNR of the GW signal.
\cWB{} has shown excellent performance in several metrics. It can analyze a large amount of data
with low computational cost.  It is sensitive to GW signals with a large variety
of waveforms.  It has been shown to be robust to calibration errors and other uncertainties,
and it also provides information about the reconstructed parameters of the signal, including an
estimate of the sky position, polarization, and waveform.  
Because $\rho$ scales with SNR, and \cWB{} attempts to search 
a very broad parameter space, even a small number of coincident, high SNR glitches can 
make high confidence detections a challenge.  
To address this, \cWB{} uses statistics based on the reconstructed noise energy
 to distinguish GW signals from glitches.
Also \cWB{} uses various search strategies to divide the parameter space and  single out specific glitch families.


The recently developed \BayesWave{} pipeline computes the Bayesian evidence for 
three competing models: the data contain only Gaussian noise, 
the data contain an astrophysical signal, or the data contain one or more glitches.
The algorithm uses a Reverse Jump Markov Chain to calculate full posterior distributions
for each of these models, and thermodynamic integration to compute the associated 
evidence.  As a detection statistic, we adopt the natural logarithm of the evidence ratio,
or ``Bayes Factor'', between the signal and the glitch model ($\log(\Baysg)$). 
The Bayes factor is the ratio of the Bayesian evidence of the signal 
hypothesis ($\Hyp_{\Sig}$) to that of the glitch hypothesis ($\Hyp_{\Gli}$):
\begin{equation}
\Baysg = \frac{p(d|\Hyp_{\Sig})}{p(d|\Hyp_{\Gli})}
\end{equation}
where $p(d|\Hyp)$ represents the marginalized likelihood that the hypothesis 
$\Hyp$ would have produced the data $d$.
BayesWave uses Markov Chain Monte Carlo methods to numerically calculate the 
evidence for each model.
Because the \BayesWave{} detection statistic is derived from a framework that expects glitches 
in the data, as opposed to assuming Gaussian noise, it may rank events in an order that better reflects the 
true probability 
that a given candidate is astrophysical in nature.  In addition, the algorithm 
calculates posterior distributions for a number of parameters, including the sky position,
central frequency, and bandwidth of any detected event.  Such information could aid in 
astrophysical interpretation.  A current limitation of \BayesWave{} is that the run time
is relatively slow, so that analyzing large data sets is impractical.



The \cWB{} detection statistic $\rho$ is derived from a ``maximum likelihood'' framework, which
calculates an optimal statistic for identifying gravitational wave bursts embedded in 
Gaussian noise.  It is natural, then, that $\rho$ scales with the signal energy present
in the data, since Gaussian noise is extremely unlikely to produce high SNR signals.
$\Baysg$, on the other hand, is calculated in a framework that directly compares a 
signal model with a glitch model.  Under this assumption, a louder event does not 
necessarily imply a larger likelihood of an astrophysical origin.  Indeed, very loud 
glitches, with SNRs of order 100, are routinely observed in the LIGO instruments, where
the bulk of astrophysical signals are most likely to be at low SNR.  Rather, the 
principal scaling for the \BayesWave{} detection statistic can be expressed as \cite{bwbfisher}:
\begin{equation}
\log\Bay_{\Sig,\Gli}\sim\mathcal O\left(N\log{\rm SNR}\right)
\label{snr_scale}
\end{equation}
where $N$ represents the number of sine-gaussian wavelets used to 
reconstruct the signal.  
This scaling can be derived analytically as the 
ratio of the \textit{Occam factors} between the two models (see \cite{bwbfisher} for details),
and emerges because real signals may be fit with both the signal and 
glitch models, but the glitch model has a much larger parameter space.
The fact that the detection statistic scales 
only with the logarithm of the SNR suggests that in this framework,
a very loud event does not provide much evidence for the signal model.
On the other hand, the detection statistic does show a strong scaling with 
the complexity of the signal in the time-frequency plane, as represented by
$N$, the number of wavelets required.  This brings out the salient feature that
distinguishes \BayesWave{} from other Burst pipelines - complexity in signal
morphology, rather than SNR, determines the significance of observed events. 

For GW signals that require only a single wavelet to reconstruct
(\textit{i.e.} sine-gaussian waveforms), $N \sim 1$, and we expect the detection
statistic to be effectively flat as a function of SNR, since 
$\log\Bay_{\Sig,\Gli}$ will scale slowly as $\log$ SNR.  On the other hand,
for a signal with a non-trivial time-frequency structure (\textit{i.e.} anything
 \textit{other than} a sine-gaussian), we expect to better resolve the signal
 with higher SNR, and so require more wavelets, so that
 \begin{equation}
N\sim1+\beta ~{\rm SNR}
\label{beta}
 \end{equation}
where $\beta$ depends only on the waveform morphology, $\beta \geq 0$, and $\beta$ is larger for waveform morphologies that have complicated time-frequency structure.




In this work, we implement a hierarchical pipeline, where \BayesWave{} 
acts as a follow-up stage, which can amplify the significance of complex GW events 
identified by \cWB.
This takes advantage of the computational efficiency and
robust trigger identification of \cWB{}, while leveraging the 
unique signal-glitch separation capabilities of \BayesWave. 
In our scheme, \cWB{} was run over 
a large set of interferometer data, represented in this study by roughly 50 days of 
data from the last science run of initial LIGO.  A nominal threshold was set for $\rho$,
and for each event above threshold, \BayesWave{} was run over 4 seconds of data centered on 
the trigger time, with the goal of calculating $\log(\Baysg)$ for the 1 second of data
around the trigger time.  This combined pipeline was run for two different testing scenarios:
\begin{itemize}
\item Binary black hole merger waveforms added to
rescaled LIGO noise, to emulate the power spectral density of expected noise in the early advanced LIGO detectors.
\item Ad-hoc waveforms used in previous Burst searches added to initial LIGO data
\end{itemize}
In the following sections, we describe these two data sets in detail, and present the results of our
testing.

\section{Testing with Binary Black Hole Waveforms}

\subsection{Data Set}

While GW signals from stellar mass black-hole mergers can be best
recovered using matched filters, intermediate mass black hole mergers
($M \sim 100 M_{\odot}$) may be well detected with burst searches,
since the time the source spends in the LIGO frequency 
band is very short.  While matched filtering may still be ideal in many
cases, burst searches present the advantage that they are robust against 
modeling uncertainties.  For example, obtaining a template bank of model 
waveforms is difficult for cases where the black holes have misaligned 
spins or are on eccentric orbits.  Previous studies have compared
the performance of burst pipelines and matched filtering pipelines in this
regime, and found them to have similar sensitivity \cite{satya2014}.

To emulate a search for intermediate mass binary black holes with Advanced LIGO, we 
used 26.6 days of coincident data from the Hanford and Livingston
LIGO detectors from the end of the last science run of initial LIGO, 
in August-October of 2010.  This data was ``recolored'' to have a 
noise power spectral density that mimics roughly what we expect from the first
science run of Advanced LIGO, with a 55 Mpc sky-averaged range for 
binary neutron star mergers \cite{obs}.  The recoloring process was intended to 
simulate noise levels from the near future detectors, while preserving 
the non-gaussian features in the data.  This was done using tools in the 
\texttt{gstlal} library \cite{gstlal} to apply filters to change the frequency
dependence of the data's power spectral density.

\subsection{Background}

\begin{figure}
\mbox{
\includegraphics[width=\columnwidth]{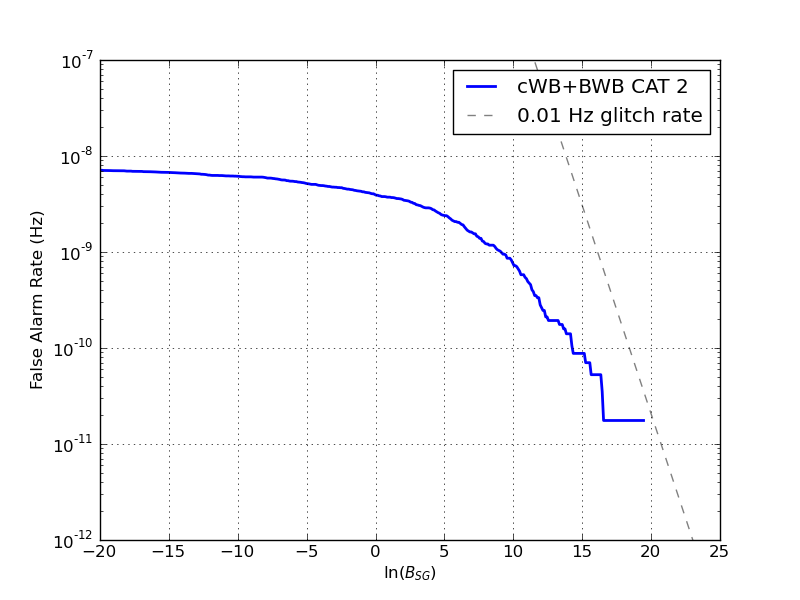} 
}
\caption{Background for the hierarchical pipeline in the the recolored data used to study IMBH signals.  The background was calculated using 1900 years of time-slide data.  The 
loudest background event has $\log \Baysg = 19$.  The gray curve shows the expected background,
based on the assumption that a glitch appears in the data every 100 seconds. }
\label{fig:bg-imbh}
\end{figure}

\begin{figure}
\mbox{
\includegraphics[width=\columnwidth]{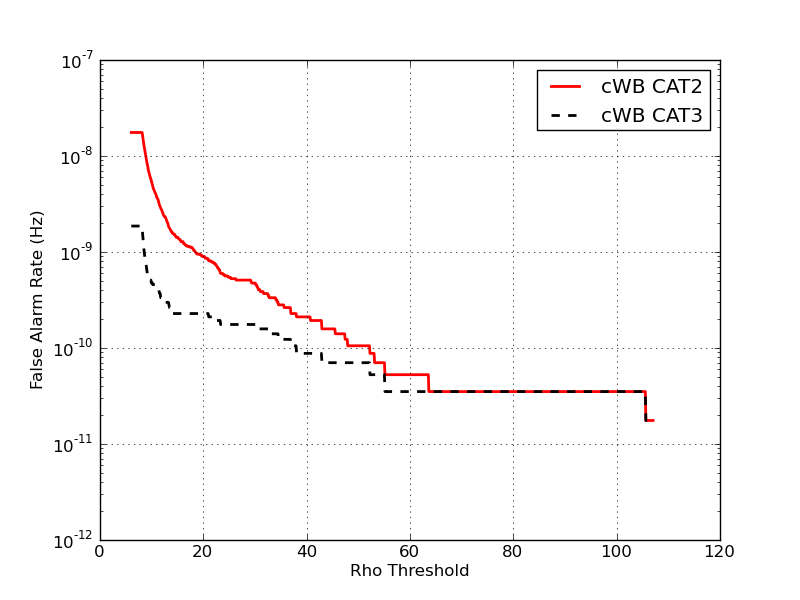} 
}
\caption{Background for the \cWB{} pipeline in the the 
recolored data used to study IMBH signals.  The background
was calculated using 1900 years of time-slide data.  The loudest 
background event has $\rho = 107$.  The two curves show the 
background after applications of ``Category 2'' (red) and ``Category 3'' 
 (black) data quality vetoes.  }
\label{fig:cwb_bg_imbh}
\end{figure}

In order to measure the rate of false-positives found by our 
hierarchical pipeline, we created 30,000 time-slide data sets by 
introducing artificial time off-sets between the Hanford and Livingston
data streams.  This background data represent 1896 years of effective livetime.  
\cWB{} was run over all of this data, searching for transients 
in a band from 16-512 Hz.
We set a nominal threshold at $\rho > 8.1$, yielding 500 background triggers 
from the time-slide data.  All 500 background triggers were processed by 
\BayesWave{} using the same bandwidth as \cWB{} to determine 
the final detection statistic, $\log(\Baysg)$, for each event.  The false alarm rate (FAR) for this 
background set is shown in Figure \ref{fig:bg-imbh}.  The loudest background event 
has $\log(\Baysg) \sim 19$.  We also 
show in Figure \ref{fig:cwb_bg_imbh} the same background set, 
as a function of the the cWB detection statistic $\rho$.  
If no additional cWB selection cuts are applied, 
the same FAR may be achieved with a threshold of 
$\rho > 107$.  We tried applying both ``Category 2'' and ``Category 3'' data 
quality cuts, as was done in initial LIGO searches \cite{s6burst}, and found that this makes 
only a small difference for the loudest several events that dominate the high SNR ``tail'' of the 
\cWB{} background (See Figure \ref{fig:cwb_bg_imbh}).  
To address the background problem, the \cWB{} trigger set would require additional processing, 
such as dividing up the parameter space, additional selection cuts, 
or a follow-up with an additional burst algorithm as described in this work.

A novel feature of the \BayesWave{} pipeline is that it allows an 
\textit{a priori} estimate of the expected background rate, which may
be compared with the results of running the hierarchical pipeline.  
Based on 
studies of LIGO noise properties \cite{s6detchar},
we expect coincident glitches in the LIGO detectors roughly every 100 seconds
($\mathrm{R_{glitch}} = 0.01$ Hz).
For glitches that mimic real signals, the Bayes Factor $\Baysg$ is dominated by the 
``Occam Factor'', an estimate of the fraction of glitch parameter space that 
is consistent with the signal model for a given event.
This immediately leads to an expectation for the background rate of glitches 
that mimic real signals by chance, 
\begin{equation}
\mathrm{FAR_{expected}} \sim  (1 / \Baysg) \times \mathrm{R_{glitch}}
\end{equation}
which 
is plotted as a gray dashed line in Figure \ref{fig:bg-imbh}.

\subsection{Injections}

To test the ability of the joint \cWB{} plus \BayesWave{} pipeline
to recover GW signals, we added two sets of simulated black hole mergers
to the data.  One set contained waveforms from binary black holes 
with component masses of 50 $M_{\odot}$, the other contained
component masses of 150 $M_{\odot}$.  Both sets were distributed 
uniformly in co-moving volume, and generated using non-spinning
effective-one-body waveforms, known as EOBNRv2 \cite{EOBNRv2}.

\begin{figure}
\mbox{
\includegraphics[width=\columnwidth]{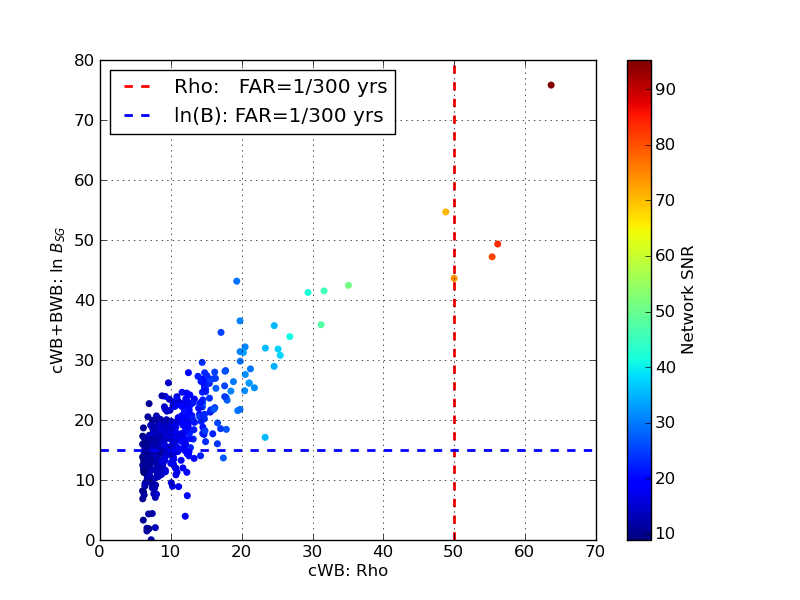} 
}
\caption{Scatter plot of two detection statistics with simulated mergers of pairs
50 solar mass black holes, comparing the hierarchical pipeline with the basic \cWB{} cuts.  
The dashed lines correspond to thresholds required for 
a false positive rate of 1 in 300 years.  Injections in the upper-left quadrant were ``promoted'' by the 
\BayesWave{} follow-up.}
\label{fig:rho_v_snr}
\end{figure}

\begin{figure}
\mbox{
\includegraphics[width=\columnwidth]{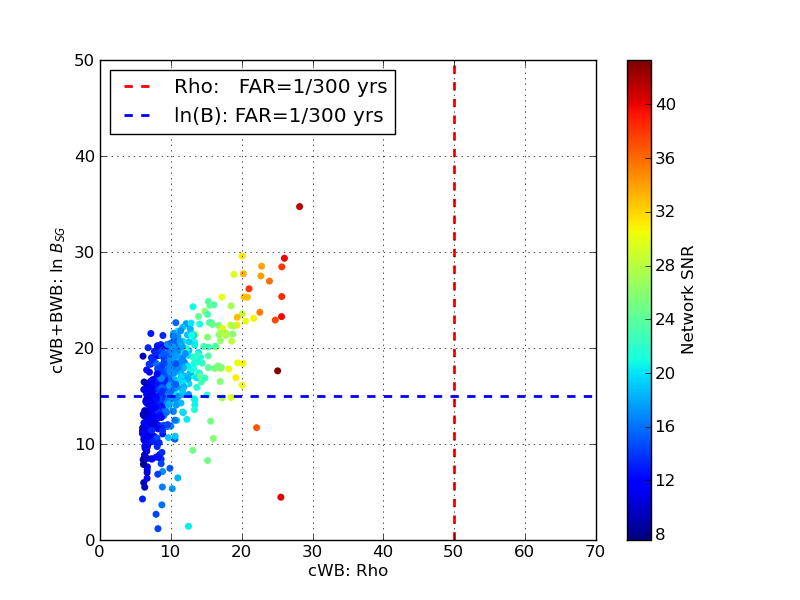} 
}
\caption{Scatter plot of two detection statistics with simulated mergers of pairs
150 solar mass black holes, comparing the hierarchical pipeline with the basic \cWB{} cuts.  
The dashed lines correspond to thresholds required for 
a false positive rate of 1 in 300 years.  Injections in the upper-left quadrant were ``promoted'' by the 
\BayesWave{} follow-up.}
\label{fig:snr_v_ev}
\end{figure}

Scatter plots of the recovered injections can be seen in Figures 
\ref{fig:rho_v_snr} and \ref{fig:snr_v_ev}.  In each figure, 
the x-axis corresponds to the \cWB{} detection statistic $\rho$,
while the y-axis shows the \BayesWave{} detection statistic 
$\log \Baysg$.  As a demonstration that the hierarchical pipeline
can make high confidence detections, the blue dashed line in each figure represents
the threshold required for a ``3-$\sigma$'' level detection in a year of observations,
corresponding to a FAR of $9 \times 10^{-11}$ Hz.  Many injections are seen to be 
well above this threshold, even with network SNR as low as $\sim 10$.  

For comparison, the red vertical line shows the $\rho$ value that would be required
for the same confidence level, if only the basic 
\cWB{} cuts were used.  The two dashed lines, then, divide the figure into
four quadrants which classify each injection as detected by the hierarchical pipeline 
but not the basic \cWB{} cuts (top left), detected by \cWB{} but not by the 
hierarchical pipeline (bottom right), detected by both approaches, or detected by neither.
The threshold required by the basic \cWB{} cuts is clearly too high to be practical - 
events would need a network SNR $>$ 50 to stand above the background.  
Any search with \cWB{}, then, would need some additional layer of processing.
The events in the top-left quadrant of Figures \ref{fig:rho_v_snr} and \ref{fig:snr_v_ev}
represent black hole merger signals which were detected with low-significance by the 
basic \cWB{}, but were ``promoted'' by the \BayesWave{} follow-up.  The fact that most
of the injections with SNR 10-50 fall in this range suggests that, for these waveforms,
the hierarchical pipeline described in this work represents a successful strategy. 
Moreover, while the blue dashed line corresponds to a 
$3 \sigma$ detection level, we note that the loudest background event in the data 
set had a value of $\log \Baysg \sim 19$.  Many of the events are seen to rank higher than this, 
suggesting that detections at even higher significance levels are possible, 
even with plausible SNR values.

\begin{figure}
\begin{center}
\mbox{
\includegraphics[width=\columnwidth]{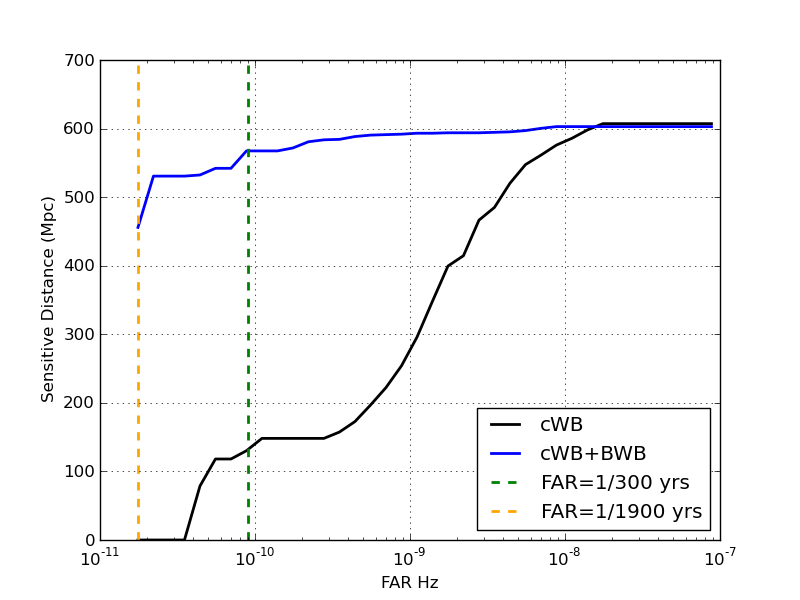} 
}
\caption{Sensitive distance for the 50-50 solar mass injections.  The sensitive distance is a measure of the effective radius to which the analysis is sensitive.
The shown curves apply only category 2 data quality cuts.}
\label{fig:dist-50}
\end{center}
\end{figure}

To quantify the performance of the hierarchical pipeline
for black hole mergers, we follow
the methodology described in \cite{satya2014} and plot 
the ``sensitive radius'' at a range of FAR values in Figure  
\ref{fig:dist-50}, a statistic that characterizes 
the effective range of the survey.  
At very high FAR values ($\sim$ 1 per 10 years), the pipeline presents no advantage 
over a basic application of \cWB{}.  However, moving to the left side of the plot, 
representing high confidence detections, the hierarchical pipeline still detects a 
large fraction of the injection set.  This means that the hierarchical pipeline is 
able to detect black hole mergers at high confidence, even in the presence of glitches,
without reliance on a matched filter technique.

\section{Testing with ad-hoc waveforms}


\subsection{Data Set}

In principle a burst search should be sensitive to a wide range of possible
signal morphologies.  In order to test this, without reliance on any particular
astrophysical models, past all-sky burst searches have made use of a suite of
ad-hoc waveforms to measure pipeline performance \cite{s6burst}.  This set includes
two extremes of waveform complexity.  At one extreme, linealy polarized sine-gaussian, 
or Morlet-Gabor, 
waveforms represent a minimum possible ``time-frequency volume'' \cite{shourov}, and so 
may be described as the simplest possible signals in this domain.  They also correspond
to the basis functions used by the \BayesWave{} pipeline, and so are best represented
by a single wavelet ($N \approx 1$).  The set also contains unpolarized 
``white-noise burst'' waveforms, which are random waveforms within a fixed 
duration and bandwidth.  In the time-frequency plane, these waveforms essentially fill
a large block with edge sizes corresponding to the duration and bandwidth of the signal.
The white noise bursts are a very poor match to the \BayesWave{} Morlet-Gabor basis, 
and so require a large number of wavelets to reconstruct.  In this sense, they 
have a very complex time-frequency structure.  The unpolarized white-noise bursts also provide 
an interesting test of the \BayesWave{} signal model, which uses an elliptical polarization 
model for GW signals.
To measure the performance of the hierarchical pipeline using these ad-hoc waveforms,  we used 51 days of coincident data from the H1-L1 network during the last science run of initial LIGO, between August and October 2010.  
The data and data quality information are both available through the LIGO Open 
Science Center \cite{losc}.  

\subsection{Background}

\begin{figure}
\mbox{
\includegraphics[width=\columnwidth]{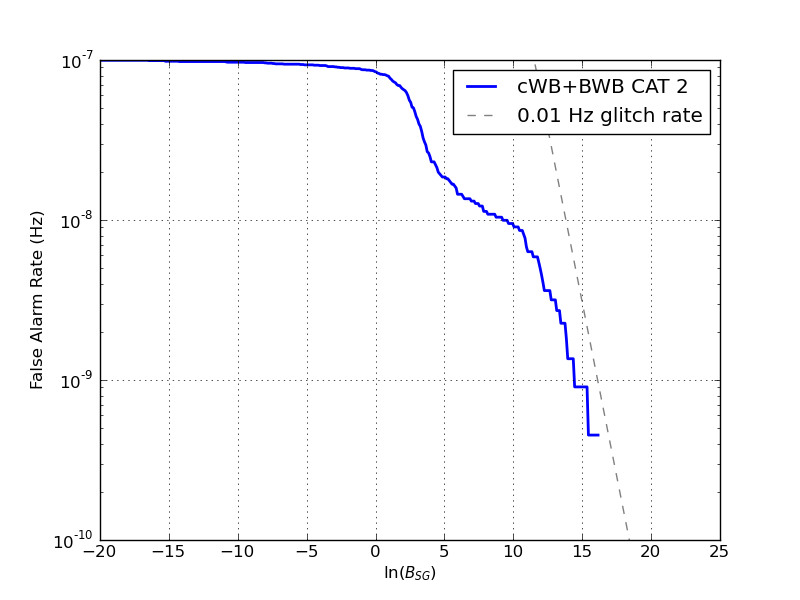} 
}
\caption{Background for the hierarchical pipeline using the LIGO data used to study ad-hoc signals.  The background was calculated using 70 years of time-slide data.  The loudest background event has $\log \Baysg = 16$.  The gray curve shows the expected background,
based on the assumption that a glitch appears in the data every 100 seconds. }
\label{fig:bg_s6}
\end{figure}

\begin{figure}
\mbox{
\includegraphics[width=\columnwidth]{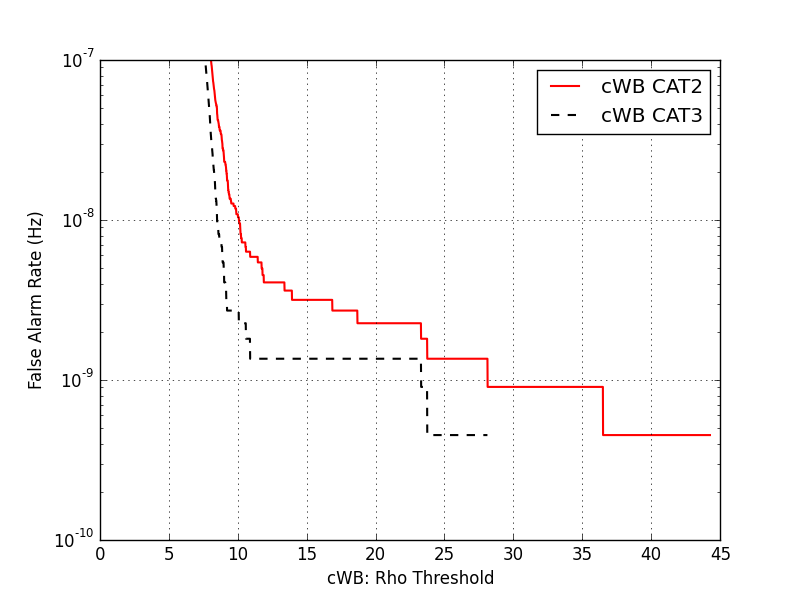} 
}
\caption{Background for the \cWB{} pipeline using the S6 
data used to study ad-hoc signals.  The background
was calculated using 70 years of time-slide data.
The two curves show the 
background after applications of ``Category 2'' (red) and ``Category 3'' 
(black) data quality vetoes.}
\label{fig:cwb_bg_s6}
\end{figure}

In order to measure the FAR of the search, the data were time-shifted
500 times to create a background set with 70 years of effective livetime.  Following 
\cite{s6burst}, 
we searched
through this data set using \cWB{} in a bandwidth from 32 to 2048 Hz.  As for the black hole 
merger data set, 
each trigger from \cWB{} above a nominal threshold ($\rho > 8$) was processed in a 
1 second window using \BayesWave{}.  In order to reduce processing time, we used
the central frequency as reported by \cWB{} to limit the bandwidth of some 
\BayesWave{} jobs.  Triggers with a central frequency less than 200 Hz were processed
with a bandwidth from 16-512 Hz, while triggers with a higher central frequency 
used a band from 16-2048 Hz.  The rate of background triggers for the 
hierarchical pipeline is shown in Figure \ref{fig:bg_s6}, and the 
corresponding FAR for the basic application of \cWB{} is shown in Figure \ref{fig:cwb_bg_s6}.

The background distributions were broadly similar to the distributions for the
recolored data set. In Figure \ref{fig:cwb_bg_s6}, we see that the basic \cWB{}
cuts lead to a ``tail'' in the distribution.   As in the IMBH data set, we see 
in Figure \ref{fig:bg_s6} that the hierarchical 
pipeline shows a distribution that is similar to expectations based on the 
known glitch rate, marked as a grey line in the figure.  For this data set, the background represents 70 years of effective
livetime, so the detection statistic of the loudest event corresponds to a FAR 
threshold of $5 \times 10^{-10}$ Hz.


\subsection{Injections: white noise bursts}

To measure the ability of the hierarchical pipeline to recover
astrophysical signals, we added simulated signals to the data set 
before running the pipeline.
Tens of thousands of injections were added to the 51 days of data.
As with the background set, this data was searched first with \cWB{}
to identify triggers with $\rho > 8$.  Unlike with the background set,
for the injections we randomly selected a sub-set
of around 200 of the \cWB{} triggers for each waveform to process with \BayesWave{}.  

\begin{figure}
\mbox{
\includegraphics[width=\columnwidth]{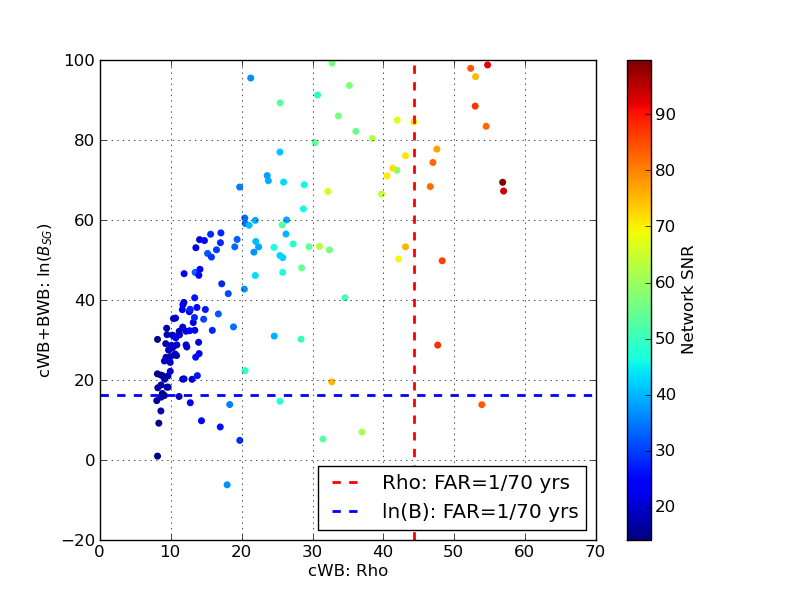} 
}
\caption{Scatter plot of white noise burst injections in the 50-150 Hz bandwidth.  
Events in the top-left quadrant were ``promoted'' to possible detections by the 
application of \BayesWave{} in the second stage of the hierarchical pipeline.}
\label{fig:wnb_scatter}
\end{figure}

We tested three different white noise burst waveforms with short duration (t $<$ 0.1 s), each with 
a different central frequency and bandwidth, as used in initial LIGO searches \cite{s6burst}.  
A fourth waveform, with longer duration and narrow bandwidth, was found incompatible 
with the \BayesWave{} parameters used in this search; the long duration ($>0.5$ s) would
require a larger data segment for power spectral density estimation and a larger maximum 
number of wavelets to cover.
An example scatter plot showing 
the results of white noise burst waveforms with a bandwidth from 50-150 Hz 
is shown in Figure \ref{fig:wnb_scatter}.  In the figure, the colorbar shows the 
network SNR, and the X and Y axes correspond to the \cWB{} and 
\BayesWave{} detection statistics, respectively.  Also shown are the loudest 
background event after category 2 data quality for both the first stage \cWB{} cuts (red, vertical line)
and the second stage \BayesWave{} ranking statistic (blue, horizontal line).  
These thresholds divide
the figure into four quadrants, representing if the event was detected at 
various stages of the hierarchical pipeline.  For example, events in the upper
left quadrant stood above the background for the hierarchical pipeline, but 
would not have been detected using only the basic version of the \cWB{} pipeline. 
We say these events were ``promoted'' by the \BayesWave{} follow-up, 
since their detection confidence increased due to this step.

Since $\rho$ scales linearly with SNR, moving from left to right across the 
figure represents growing SNR.  As the network SNR grows from 10 to 30, 
$\log \Baysg$ is seen to grow quickly.  This is expected for complex waveforms:
signals with higher SNR require more wavelets
to reconstruct, as in Equations \ref{snr_scale} and \ref{beta}.  Figure
\ref{fig:wnb_scatter} shows that the typical event with network SNR $>$ 15
has a bayes factor that exceeds the loudest background event, and so could 
be detected with high confidence.  To reproduce a result like this 
was not possible using only a basic application of \cWB{}; \BayesWave{} or 
some other form of follow-up was required for high confidence detections.

\begin{figure}
\mbox{
\includegraphics[width=\columnwidth]{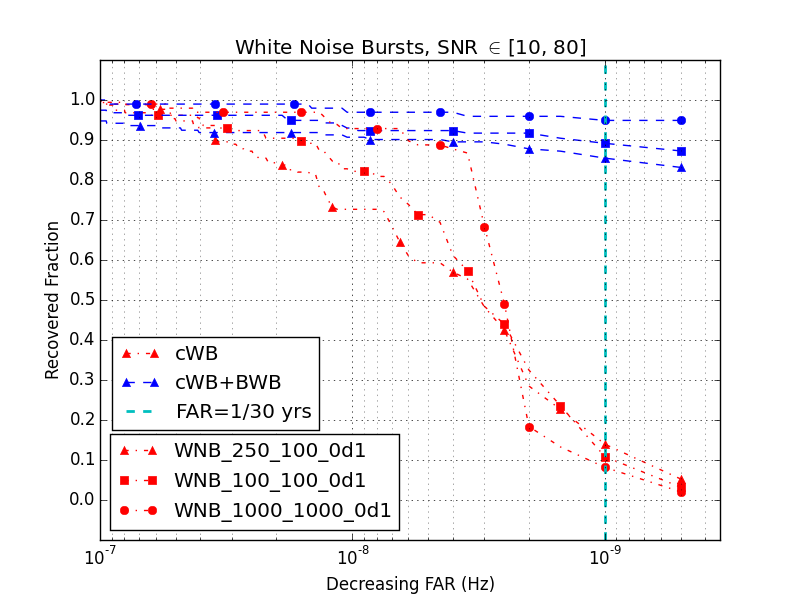} 
}
\caption{Fraction of white noise burst injections identified in the first stage
which were recovered at various FAR thresholds, after applying CAT 2 vetoes.
The right side of the figure indicates higher confidence detections.  The hierarchical pipeline
performs well recovering complex waveforms at high confidence.
For waveform details, see \cite{s6burst}.}
\label{fig:wnb_eff}
\end{figure}

To quantify the performance of the second stage in the hierarchical pipeline, 
Figure \ref{fig:wnb_eff} shows the performance of both the hierarchical pipeline
and the first stage \cWB{} cuts at various FAR thresholds.  The efficiency
shown on the Y-axis uses the number of triggers identified by \cWB{} with network 
SNR $<$ 80 as the denominator, and shows the fraction of these events recovered 
above the threshold on the X-axis.  The efficiency of recovering events at 
high confidence is seen on the right side of the plot.  The second stage of 
the pipeline ``promotes'' 80-95\% of the events to a FAR $<$ 1/70 years, where essentially
none of the events in this SNR range cross this threshold using only the 
first stage of the pipeline.

\subsection{Injections: Sine-gaussians}

We also tested a variety of sine-gaussian waveforms with different 
central frequencies and quality factors.  These waveforms match the 
basis wavelets used by \BayesWave{}, so they have minimal complexity, 
and in Equation \ref{snr_scale}, $N \approx 1$ ($\beta = 0$).  This means that
$\log \Baysg$ scales with the logarithm of SNR.  This leads to
a counter-intuitive conclusion: The most challenging waveforms for 
\BayesWave{} to detect are sine-gaussians, because they match the basis used
by the pipeline.  Potentially, future versions of the code could target these
signals by using a more precisely formulated glitch model, though at some point
one would encounter the basic problem that some LIGO glitches do, in fact, 
mimic sine-gaussian signals.

\begin{figure}
\mbox{
\includegraphics[width=\columnwidth]{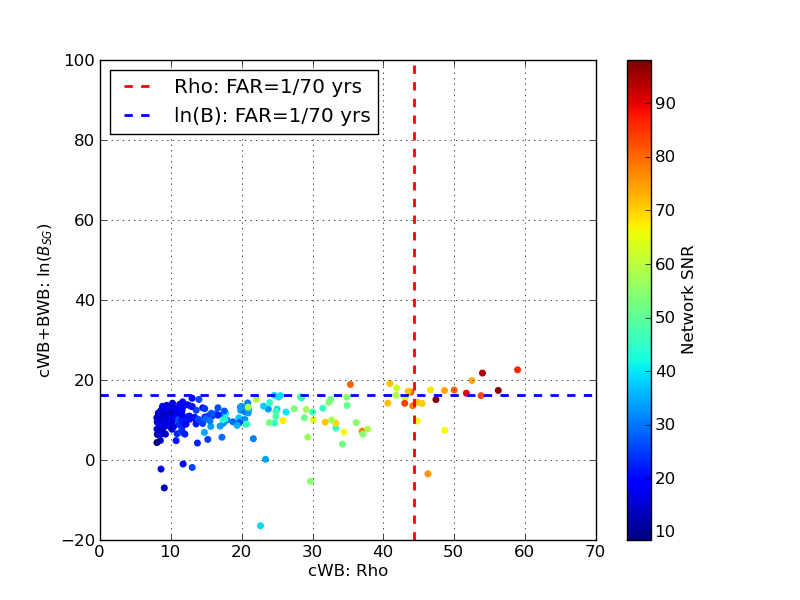} 
}
\caption{Scatter plot of 153 Hz, Q9 sine-gaussian injections.  
The injections have a simple time-frequency structure, so the Bayes Factor scales
only weakly with SNR.}
\label{fig:sg_scatter}
\end{figure}

An example of this logarithmic scaling may be seen in Figure \ref{fig:sg_scatter}.
Compared with the white noise bursts in Figure \ref{fig:wnb_scatter}, the 
Bayes Factor of the sine-gaussian injections grows very slowly with SNR.
The result is that nearly all of the events have $\log \Baysg$ in the range 
$5-20$.  Comparing with the background distribution, we can see this means
the hierarchical pipeline will detect most sine-gaussian signals with a FAR
$10^{-10} - 10^{-8}$ Hz, or about 1 background event per 3-300 years.
This FAR is too low to make first detections, though it may be in the right range
for identifying interesting candidates, and may be appropriate in a future
scenario where GW detections are common.  

\begin{figure}
\mbox{
\includegraphics[width=\columnwidth]{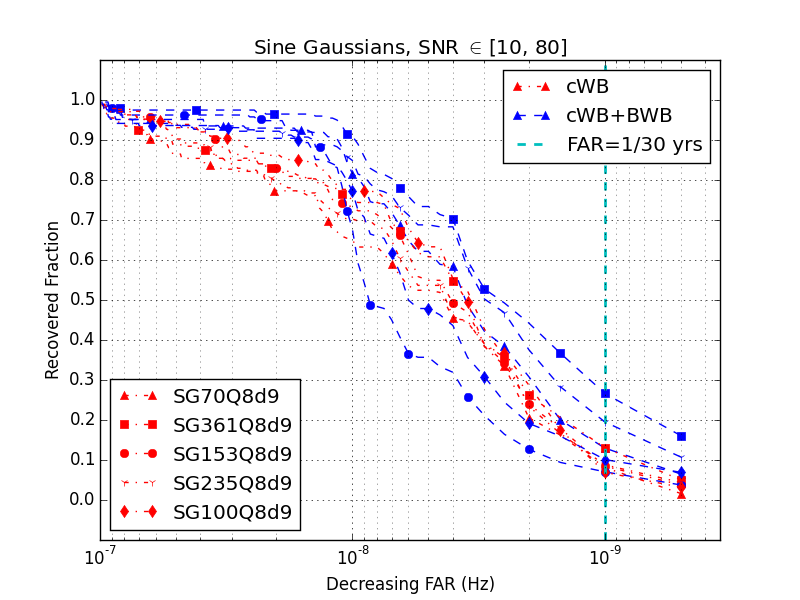} 
}
\caption{Fraction of sine-gaussian injections detected at various FAR thresholds.
The simple waveforms are detected at low confidence, but not at high confidence. For waveform details, see \cite{s6burst}. 
}
\label{fig:sg_eff}
\end{figure}

The FAR range where \BayesWave{} detects sine-gaussian signals loosely corresponds to 
the FAR levels where we see long tails in the naive \cWB{} background.
Moreover, experience with \cWB{} suggests that typical glitch populations 
that pass basic selection cuts often have a simple time-frequency structure.
The statistical framework developed for \BayesWave{} predicts this effect:
glitches with simple time-frequency structure may appear as coincident in 
both detectors, and so mimic real signals.  However, glitches with complex 
time-frequency structure are highly unlikely to appear identical in two or 
more detectors, and so can typically be rejected by testing for signal power 
inconsistent with the astrophysical model.  One could argue the
\BayesWave{} ranking by signal complexity is a natural approach for this 
reason: signals with simple time-frequency structure may be plausibly
explained as a glitch, while signals with complex time-frequency structure 
are extremely unlikely to appear consistent with a GW signal in two detectors by chance.  
The conclusion, then, is that
short duration sine-gaussian waveforms represent a special case which are challenging to detect 
at high confidence due to similarity with detector glitches.  
The results in Figure \ref{fig:sg_eff} show that the hierarchical pipeline and basic \cWB{} perform 
at a similar level for these waveforms.  
If nature indeed produces simple waveforms, from sources like cosmic string cusps, very high mass binary black holes, etc.,
perhaps they can be recovered with more specialized burst searches,
or populations of such events could stand
above the background.

\section{Discussion}

In this work, we introduced a hierarchical pipeline which combines 
two previously described algorithms for finding GW burst signals.
The performance of the hierarchical pipeline was tested using a variety of 
simulated waveforms
embedded in data from the two detector LIGO network.  The testing 
demonstrated that for this data set:
\begin{itemize}
\item For complex waveforms requiring several wavelets to fit, including white noise 
bursts and binary black hole merger signals,  the hierarchical pipeline can make 
high confidence detection for low SNR events.

\item For simple waveforms (for example, sine-gaussians), the hierarchical 
pipeline does not improve the detection confidence and other alternative 
approaches need to be explored.

\item The distribution of background events studied with \BayesWave{} was broadly consistent
with a simple, predictive model.

\end{itemize}

The implication for the early Advanced LIGO network is clear: using a 
detection statistic that accounts for waveform complexity,
as was done here with \BayesWave{}, enables high-confidence
detection of short GW bursts even in the presence of loud glitches.

The fact that \cWB{} uses a detection 
statistic that scales linearly with SNR means that even a 
single loud glitch in the background set requires special attention
to enable high-confidence detections. The \cWB{} background typically 
contains large ``tails'', and 
so follow-up or specialized cuts are required.  On the other hand, 
glitches which contain significant time-frequency structure,
and so require multiple wavelets to reconstruct, are 
extremely unlikely to have the same time-frequency structure
in two detectors.  \BayesWave{} leverages this feature of the 
data to assign a high detection confidence to signals with 
complex time-frequency structure, and a low confidence to simple 
signals.  This means that simple, loud glitches are ``down-weighted''
by \BayesWave{}, while complex glitches are most likely rejected 
due to a lack of coherence between detectors.  The fact that 
\BayesWave{} uses this important morphology information to 
rank events, while \cWB{} ranks mainly by coherent SNR, 
are complimentary features of the two algorithms.



Finally, we note that the strong performance shown here is not necessarily 
restricted to this particular implementation.  Rather, this is a 
result of a detection statistic that better reflects the properties
of LIGO data than SNR based schemes.  Such a statistic could be 
implemented in a less computationally intensive framework, so that 
a single stage pipeline may show similar performance.  In
this sense, we hope that this work marks a turning point in the culture
of GW transient searches, so that our community can move beyond
only looking for the very loudest signals, and instead give proper statistical
weight to waveform morphology in our searches.



\section{Acknowledgements}

Thank you to Kent Blackburn,  Reed Essick, Tjonnie  Li, Joey Shapiro Key,  Patricia
Schmidt,  Tiffany Summerscales, Michele  Vallisneri, Salvatore Vitale, Leslie Wade, and 
Alan Weinstein  for  helpful  
conversations  about  this  work.  Thanks also to Seth Kimbrell and Francesco 
Pannarale for contributions to \BayesWave{} development.
LIGO was constructed by the California Institute of Technology and Massachusetts Institute of
Technology with funding from the National Science Foundation and operates under cooperative 
agreement PHY-0757058 . T.B.L. acknowledges the support of NSF LIGO grant, award PHY-1307020.  
This paper carries LIGO Document Number LIGO-P1500137-v7.

\bibliography{bayeswave_iii.bib}

\end{document}